\documentclass[prd,aps,preprint,amsmath,nofootinbib,amssymb,eqsecnum,showkeys,tightenlines]{revtex4-1}
\usepackage{hyperref} % should be commented out if the tex file will be compiled with latex in arXiv!!! (pdflatex is fine)
\usepackage{slashed}
\usepackage{epsfig,latexsym,cancel,amssymb,amsmath,verbatim,mathrsfs}
\usepackage{color}
\usepackage{graphicx}
\usepackage{caption}
\usepackage{subcaption}
\usepackage[inline]{enumitem}
\usepackage[multidot]{grffile}  % allow figure file names to include dots
\usepackage{dcolumn}
\usepackage{bm}
\usepackage{amsmath}
\usepackage{amsfonts}
\usepackage{amssymb}
\usepackage{color}
\usepackage{latexsym}
\usepackage{slashed} % slash mark
\usepackage{pstricks}
\usepackage{indentfirst}
\usepackage{mathrsfs}
\usepackage{multirow}
\usepackage{epsfig,psfrag}
\usepackage{setspace} % spacing
\usepackage[utf8]{inputenc} % accept utf-8 input encoding
\usepackage[scientific-notation=true]{siunitx} % comprehensive units

\allowdisplaybreaks % allow eqnarray breaks
\newcommand{\beq}{\begin{eqnarray}}% can be used as {equation} or  {eqnarray}
\newcommand{\eeq}{\end{eqnarray}}

\graphicspath{{fig/}}

\allowdisplaybreaks

\begin{document}

%%%%%%%%%%%%%%TITLE AFFILIATIONS ETC%%%%%%%%%%%%%%%%%%%%%%%%%%%%%%%%%%%%%%%%%%%%%%%%%%%%%%%%%%%%

\title{\texorpdfstring{\Large  Cancellation mechanism of dark matter direct detection in Higgs-portal and vector-portal models}{}}

\author{Chengfeng {\sc Cai}\footnote{These authors contributed equally to this paper.}}
%\email{caichf3@mail.sysu.edu.cn}
\author{Yu-Pan {\sc Zeng}\footnote{These authors contributed equally to this paper.}}
\affiliation{School of Physics, Sun Yat-Sen University, Guangzhou 510275, China}
\author{Hong-Hao {\sc Zhang}}
\email[Corresponding author. ]{zhh98@mail.sysu.edu.cn}
\affiliation{School of Physics, Sun Yat-Sen University, Guangzhou 510275, China}
%%%%%%%%%%%%%%%%%%%%%%%%%%%%%%%%%%%%%%%%%%%%%%%%%%%%%%%%%%%%%%%%%%%%%%%%%%

%%%%%%%%%%%%%%%%%%%%%%%%%%%%%%%%%%%%%%%%%%%%%%%%%%%%%%
%%%%%%%%%%%%%%%%%%%%%%%%%%%%%%%%%%%%%%%%%%%%%%%%%%%%%%
\begin{abstract}
We present two alternative proofs for the cancellation mechanism in the U(1) symmetric pseudo-Nambu-Goldstone-Boson Dark Matter (pNGB DM) model. They help us to have a better understanding of the mechanism from multi-angle, and inspire us to propose some interesting generalizations. In the first proof, we revisit the non-linear representation method and rephrase the argument with the interaction eigenstates. In this picture, the phase mode (DM) can only have a trilinear interaction  with a derivative-squared acting on the radial mode when the DM is on-shell. Thus, the DM-quark scattering generated by a mass mixing between the radial mode and the Higgs boson vanishes in the limit of zero-momentum transfer. Using the same method, we can easily generalize the model to an SO(N) model with general soft-breaking structures. In particular, we study the soft-breaking cubic terms and identify those terms which preserve the cancellation mechanism for the DM candidate. In our discussion of the second method, we find that the cancellation relies on the special structure of mass terms and interactions of the mediators. This condition can be straightforwardly generalized to the vector-portal models. We provide two examples of the vector-portal case where the first one is an $\textrm{SU(2)}_{L}\times \textrm{U(1)}_Y\times \textrm{U(1)}_X$ model and the second one is an $\textrm{SU(2)}_{L}\times \textrm{U(1)}_Y\times \textrm{U(1)}_{B-L}\times \textrm{U(1)}_X$ model. In the first model the vector mediators are the $Z_\mu$ boson and a new U(1)$_X$ gauge boson $X_\nu$, while in the second model the mediators are the $\textrm{U(1)}_{B-L}$ and  $\textrm{U(1)}_{X}$ gauge bosons. The cancellation mechanism works in both models when there are no generic kinetic mixing terms for the gauge bosons. Once the generic kinetic mixing terms are included, the first model requires a fine-tuning of the mixing parameter to avoid the stringent direct detection bound, while the second model can naturally circumvent it.
	\\[.3cm]
\end{abstract}
\maketitle
\newpage

\section{Introduction}
Cosmological and astrophysical observations indicate that the energy of the universe consists of substantial cold Dark Matter (DM)~\cite{Planck:2018vyg}, which cannot be explained by the Standard Model (SM) of particle physics. By far the most attractive candidate of DM is the Weakly Interacting Massive Particles (WIMPs), which couple to the SM particles with a strength similar to the weak interaction. The WIMP models are interesting not only because they can naturally explain the data of DM relic abundance by the thermal production mechanism, but also because they may be detected in terrestrial experiments. In recent years, there are many underground dark matter direct detection experiments, e.g., XENON1T~\cite{Aprile:2018dbl}, LUX~\cite{Akerib:2016vxi}, and PandaX-4T~\cite{PandaX-4T:2021bab}, searching for signals of DM-nuclei scattering. However, there is still a null result from all these experiments even the detection sensitivity has been improved by successive upgrades. The absence of direct detection signal can be explained by a super weak interaction between the dark sector and the SM, but then it is hard to obtain the observed relic density of DM by the well-studied freeze-out production framework.

In recent years, the pseudo-Nambu-Goldstone-Boson (pNGB) dark matter models, which naturally predict a suppressed direct detection signal, have drawn much attention. The model was firstly established in Ref.\cite{Gross:2017dan}, where a cancellation mechanism of Higgs portal DM-nuclei scattering is found in their template model. The cancellation mechanism is based on a soft broken global U(1) symmetry and the pNGB property of the DM candidate. It is found that DM-nuclei scattering processes happen in the t-channel mediated by two Higgs bosons and their amplitudes cancel automatically in the zero-momentum transfer limit. On the other hand, since the DM pair annihilation processes are mainly through the s-channel which is not suppressed in general, the correct relic density can be easily achieved as in the usual WIMPs scheme. Because of this nice property, many works have followed up the model and studied it in different aspects. For example, many papers have discussed various phenomenologies of the model~\cite{Glaus:2020ihj,Arina:2019tib,Scaffidi:2020nti,Ruhdorfer:2019utl,Cline:2019okt,Kannike:2019wsn,Huitu:2018gbc,Ishiwata:2018sdi,Azevedo:2018exj,Sanderson:2018lmj,Azevedo:2018oxv,Cheng:2018ajh,Lebedev:2017uyk,Zeng:2021moz,Claude:2021sye,Coito:2021fgo,Abe:2021jcz,Abe:2021vat}. In Ref.\cite{Abe:2020iph,Okada:2020zxo,Abe:2021byq,Okada:2021qmi}, the UV completion and its impact on phenomenology have been studied. The cancellation mechanism is also generalized to 2HDM+pNGB DM~\cite{Jiang:2019soj,Zhang:2021alu,Biekotter:2021ovi}, O(N)/O(N-1) model~\cite{Alanne:2018zjm}, and SU(N) model~\cite{Karamitros:2019ewv}.

In this work, we are going to revisit the pNGB DM model and discuss two simple methods for proving the cancellation mechanism. In the first proof, we will revisit the non-linear representation which has been considered in Ref.~\cite{Alanne:2018zjm}. We will rephrase the argument in a way that the cancellation becomes obvious. \footnote{In Ref.\cite{Ruhdorfer:2019utl}, the non-linear representation of $S$ was also mentioned and a dim-6 derivative operator was obtained by integrating the heavy radial mode. In our proof, we will keep the radial mode. The same proof was also presented in Ref.\cite{Lebedev:2021xey}. } Using the non-linear representation can help us to generalize the model in different ways. For example, extending the symmetry to SO(N) is straightforward in this picture. In Ref.\cite{Alanne:2018zjm}, the SO(N) with masses degenerate pNGBs has been studied, so we will focus on  more general soft-breaking structures including non-degenerate spectrum and also the soft-breaking cubic terms. We find that with certain conditions, some cubic terms can preserve the cancellation for the DM candidate.

In the second proof, we use the linear representation and show that the combination of the CP-even scalars coupling to DM can be redefined as a new scalar boson with no mass mixing to the SM Higgs boson. Instead, a kinetic mixing between them is generated and it is the only portal connecting SM fermions to the DM. The suppression of the cross section is then caused by the fact that the kinetic mixing terms of the mediators are effectively negligible in the t-channel DM-quarks scattering processes.

Inspired by the second proof, we generalize the cancellation mechanism to the vector-portal cases. It is well known that if the DM is a Dirac fermion (or complex scalar) which couples to the gauge boson, the gauge boson mediated DM-nucleon scattering cross section will be too large to accommodate the current direct detection bounds. Therefore, finding a mechanism that can naturally generate a small direct detection cross section without suppressing the DM annihilation cross section is phenomenologically interesting. We will establish two template models to illustrate how the mechanism works. In the first model, we introduce a new gauge boson from a U(1)$_X$ symmetry and show that the DM-quark scattering mediated by $Z_\mu$ and $Z'_\mu$ bosons cancel in the zero-momentum transfer limit. However, the cancellation is violated if there is a generic kinetic mixing between the U(1)$_Y$ and the U(1)$_X$ gauge bosons. In our second model, we propose a $\mathrm{U(1)}_{B-L}\times \mathrm{U(1)}_X$ extension in which cancellation occurs between the two new gauge bosons. We find that, even if the generic kinetic mixing term violates the cancellation, the direct detection bound can still be circumvented since the mixing can be naturally small in this case if it originates from 2-loop corrections.

This paper is organized as follows. In section \ref{sect.proof}, we establish our two proofs of the cancellation mechanism for the Higgs-portal model and introduce the SO(N) generalization. In section \ref{sect.vec}, we discuss the cancellation mechanism for vector-portal models and provide two examples. Finally, our conclusions are given in section \ref{concl.}.
For the reader's convenience, we briefly review the original proof of the cancellation mechanism given in Ref.~\cite{Gross:2017dan} in Appendix \ref{app1}. In Appendix \ref{app2}, we briefly introduce a UV origin of a spurion field, $\kappa_{ijk}$, which is presented in section \ref{sect.proof}.

\section{Two proofs for the cancellation mechanism}\label{sect.proof}
\begin{figure}[tb]
	\begin{subfigure}[b]{.3\textwidth}
		\centering
		\includegraphics[width=\textwidth]{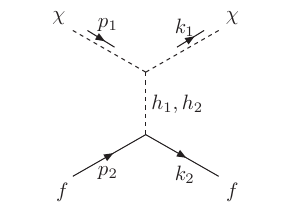}
		\caption[]{}\label{fig1a}
	\end{subfigure}\hfill
	\begin{subfigure}[b]{.3\textwidth}
	\centering
	\includegraphics[width=\textwidth]{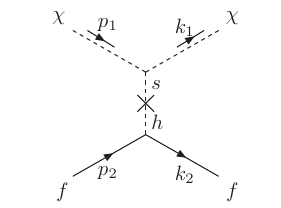}
	\caption{}\label{fig1b}
\end{subfigure}\hfill
	\begin{subfigure}[b]{.3\textwidth}
	\centering
	\includegraphics[width=\textwidth]{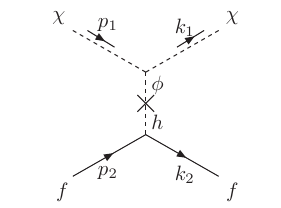}
	\caption{}\label{fig1c}
\end{subfigure}
	\caption{The three different points of view for the same DM-$f$ (SM fermions) scattering process. Plot (a) represents two diagrams with different mass-eigenstate mediators which give rise to  a miraculous cancellation. Plot (b) is the diagrammatic representation of our first proof, while Plot (c) is the perspective of our second proof.}
	\label{feyndiag1}
\end{figure}

First of all, let us briefly review the basic ideas proposed in Ref.~\cite{Gross:2017dan}. The model proposed therein consists of the SM plus an extension with a complex scalar $S$. A global U(1) symmetry of $S$ is spontaneously broken by its non-zero vacuum expectation value (VEV), and thus a Goldstone boson emerges. The Goldstone boson acquires a mass if the U(1) symmetry is softly broken. To be precise, the potential of the SM Higgs field $H$ and the scalar $S$ is given by
\beq\label{V1}
V&=&-\mu^2|H|^2-\mu_S^2|S|^2+\lambda|H|^4+\lambda_S|S|^4+2\lambda_{SH}|H|^2|S|^2\nonumber\\
&&-\frac{\mu_S^{\prime2}}{4}(S^2+S^{\ast 2}).
\eeq
The first line of eq.\eqref{V1} respects the U(1) symmetry, while the second line is a soft-breaking term. The SM gauge symmetry and the global U(1) symmetry are spontaneously broken by the VEVs of $H$ and $S$, respectively. In the unitary gauge, we can write $H$ and $S$ as
\beq\label{linear}
H=\begin{pmatrix}0\\ \frac{v+h}{\sqrt{2}}\end{pmatrix},\quad S=\frac{v_s+s+i\chi}{\sqrt{2}}~.
\eeq
The CP-even component $s$ will mix with $h$, while the CP-odd component $\chi$ is a pseudo-Nambu-Goldstone Boson (pNGB) playing the role of a DM candidate. The masses of $(h,~s)$ and $\chi$ can be easily obtained by finding the stationary point of the potential and the results are shown as follows
\beq
M_{even}^2=\begin{pmatrix}2\lambda v^2&2\lambda_{SH}vv_s\\ 2\lambda_{SH}vv_s&2\lambda_Sv_s^2\end{pmatrix}, \qquad m_\chi&=&\mu_S^\prime~.
\eeq
The mass matrix of $(h,~s)$ can be diagonalized by an orthogonal matrix $O$ as $M_{diag}^2=OM_{even}^2O^T$, while the mass eigenstate is $(h_1,~h_2)=(h,~s)O^T$. The $\chi$-quark scattering is through the t-channel mediated by $h_1$ and $h_2$ as shown in FIG.\ref{fig1a}. It has been proved by Ref.\cite{Gross:2017dan} that the amplitude of this process vanishes in the zero-momentum transfer limit ($t\to0$), since there is a cancellation between the two diagrams corresponding to the $h_1$ and $h_2$ mediators.

Although the $\chi$-quark scattering is suppressed due to the cancellation, the annihilation cross section for $\chi$ pairs is not suppressed since $\chi+\chi\to \bar{f}+f$ is through the s-channel and the $s$ variable is not necessarily small compared to the masses of the mediators. Therefore this model can easily fit the relic density data and avoid the stringent direct detection bound at the same time.

In the following subsections, we are going to introduce two methods to prove the cancellation mechanism for DM-quark scattering and discuss some possible generalizations.
\subsection{The first proof}
In this subsection, we revisit the non-linear representation proof of the cancellation mechanism in the U(1) model~\cite{Ruhdorfer:2019utl}, and rephrase it in a more simpler way~\footnote{This proof is also mentioned in Ref.\cite{Lebedev:2021xey}.}. This leads us to a better understanding of the cancellation mechanism, and helps us to generalize the model. In the non-linear representation, the complex singlet $S$ is written as
\beq\label{nonlinear_reps}
S=\frac{v_s+s}{\sqrt{2}}e^{i\frac{\chi}{v_s}}~,
\eeq
where $s$ is a scalar that mixes with the $h$, while $\chi$ is the DM candidate.
Substituting eq.\eqref{nonlinear_reps} into the potential, we find that the only terms involving $\chi$ come from the soft-breaking terms:
\beq
V_{soft}&=&-\frac{\mu_S^{\prime2}}{4}(v_{s}+s)^2\cos\left(\frac{2\chi}{v_{s}}\right)\nonumber\\
&=&-\frac{\mu_S^{\prime2}}{4}v_s^2\left(1+\frac{2s}{v_{s}}+\frac{s^2}{v_{s}^2}\right)\left(1-\frac{2\chi^2}{v_{s}^2}+...\right),
\eeq
where we have expanded the cosine function up to order $\chi^2$ in the second line.
We read off the mass squared of the pNGB $\chi$ as
\beq
m_{\chi}^2=\mu_S^{\prime2}~,
\eeq
and find that there is a $s\chi^2$ trilinear coupling,
\beq\label{tri1}
\mathcal{L}_{s\chi^2}^{(1)}=-\frac{m_{\chi}^2}{v_{s}}s\chi^2~,
\eeq
arising from the potential. In the non-linear picture, the kinetic term of $S$ consists not only of the kinetic terms of $s$ and $\chi$, but also includes derivative interactions. To be precise, the kinetic term is
\beq
\mathcal{L}_{kin}&=&(\partial_\mu S)^\ast\partial^\mu S\nonumber\\
&=&\frac{1}{2}(\partial_\mu s)^2+\frac{1}{2}(\partial_\mu \chi)^2+\frac{s}{v_{s}}(\partial_\mu \chi)^2+\frac{1}{2}\frac{s^2}{v_{s}^2}(\partial_\mu \chi)^2~.
\eeq
We see that the third term is another source of $s\chi^2$ trilinear coupling, which can be rewritten it in equivalent form as
\beq\label{tri2}
\mathcal{L}_{s\chi^2}^{(2)}&=&\frac{s}{v_{s}}(\partial_\mu \chi)^2\nonumber\\
&=&\frac{1}{v_{s}}\left\{\partial_\mu\left[s \chi\partial^\mu \chi-\frac{1}{2}(\partial^\mu s) \chi^2\right]+\frac{1}{2}(\partial^2s)\chi^2-s\chi\partial^2\chi\right\}~.
\eeq
The first term is a total derivative and can be dropped in the action. Combining eqs.\eqref{tri1} and \eqref{tri2}, the full $s\chi^2$ trilinear interaction is
\beq
\mathcal{L}_{s\chi^2}=\frac{1}{2v_{s}}(\partial^2s)\chi^2-\frac{s}{v_{s}}\chi(\partial^2+m_{\chi}^2)\chi~.
\eeq
The first term contributes a coupling proportional to the momentum squared of the $s$ field, while the second term vanishes when $\chi$ is on-shell. To the tree level, the $\chi$-quark scattering can only be mediated by the $s$ boson which mixes with the Higgs field $h$ (see FIG.\ref{fig1b}). We expect that the amplitude must be proportional to  $t=(p_1-k_1)^2$ as
\beq
i\mathcal{M}\sim\left(\frac{-it}{v_s}\right)\frac{i}{t-m_s^2}\left(-i2\lambda_{SH}vv_s\right)\frac{i}{t-m_h^2}\left(\frac{-im_f}{v}\right)\bar{u}_f(k_2)u_f(p_2)~,
\eeq
which vanishes when $t\to0$.

Note that, if we include following soft breaking terms,
\beq\label{U1cubicsoft}
V'_{soft}&=&-\kappa_1^3(S+S^\ast)-\kappa_2|S|^2(S+S^\ast)-\kappa_3(S^3+(S^\ast)^3)\nonumber\\
&=&\frac{1}{2}\left[\frac{\sqrt{2}\kappa_1^3}{v_s}+\frac{\sqrt{2}}{2}(\kappa_2+9\kappa_3)v_s\right]\chi^2+\frac{1}{2}\left[\frac{\sqrt{2}\kappa_1^3}{2v_s}+\frac{3\sqrt{2}}{4}(\kappa_2+9\kappa_3)v_s\right]\frac{2s}{v_s}\chi^2+\dots,\nonumber\\
\eeq
which involving odd numbers of $S$, the cancellation property is not preserved in general, unless $\kappa_{1,2,3}$ satisfy~\footnote{In ref.\cite{Alanne:2020jwx}, the cancellation condition for the cubic terms had been discussed.
\beq\label{speccond}
\kappa_1^3=\frac{1}{2}(\kappa_2+9\kappa_3)v_s^2~.
\eeq
}
 However, the linear term $S+S^\ast$, the cubic terms $|S|^2(S+S^\ast)$, and $S^3+(S^{\ast})^3$ usually have different origins. The $|S|^2S$ operator is a cubic term generated by a spurion with 1 unit of U(1) charge, while the $S^3$ operator is a cubic term generated by a spurion with 3 units of charge. Although the linear term and the $|S|^2S$ term has the same charge, their cancellation requires a special relation between $\kappa_1$ and $\kappa_2$ as $\kappa_{1}^3=\kappa_2v_s^2/2$, if $\kappa_3$ is set to be $0$. This is a blind-spot of direct detection due to an accidental relation between two unrelated parameters, however, it is not satisfied in a general case. There is no symmetry or fundamental principle to guarantee the relation \eqref{speccond}. Therefore, if we want the cancellation property to be obtained automatically, it is better to forbid the linear and cubic terms by assuming a $Z_2$ symmetry with $S\to-S$.

 There is a more general soft-breaking potential which includes a term as $\kappa_{HS}|H|^2(S+S^\ast)$, but it can be removed if we shift $S$ by a constant $S\to S+s_0$, where $s_0=-\kappa_{HS}/2\lambda_{SH}$. The couplings, $\mu^2,~\mu_S^2,~\mu_{S}^{\prime2},~\kappa_1^3$, also change respectively, and the cancellation conditions, in terms of the new couplings, are the same as the discussion above.

The non-linear representation is useful for generalizing this model. We easily see that the cancellation mechanism still works if the model is extended with more Higgs doublets which are neutral with respect to the global U(1). For example, the 2HDM~\cite{Jiang:2019soj} and NHDM extensions do not violate the cancellation since these Higgs fields only couple to the radial mode $s$. A more interesting generalization is to consider a global SO(N) symmetry which is spontaneously broken to SO(N-1). The SO(N) is also softly broken so that the Goldstone bosons can acquire masses.

We will consider an SO(N) model consisting of a real scalar field $\Phi$ which is in the fundamental representation of the SO(N) group. To simplify things as much as possible at the beginning, we first consider a Lagrangian without soft-breaking cubic terms,
\beq
\mathcal{L}&=&(D_\mu H)^\dag D^\mu H+\mu^2|H|^2-\lambda|H|^4\nonumber\\
&&+(\partial_\mu \Phi)^T \partial^\mu\Phi-\Phi^TM^2\Phi-\lambda_\Phi(\Phi^T\Phi)^2-2\lambda_{H\Phi}|H|^2(\Phi^T\Phi)
\eeq
where $M^2$ is an arbitrary symmetric N$\times$N real matrix. We can rotate $\Phi$ by an orthogonal transformation to a convenient basis, $\hat{\Phi}=O\Phi$, such that the mass matrix $\tilde{\mu}^2=OM^2O^T$ is diagonal. Assuming that the diagonal matrix $\hat{\mu}^2$ takes the form
\beq\label{diagmass}
\tilde{\mu}^2=\mathrm{diag}\{m_1^2-\mu_\phi^2,m_2^2-\mu_\phi^2,...,m_i^2-\mu_\phi^2,...,m_{N-1}^2-\mu_\phi^2,-\mu_\phi^2\},
\eeq
where all the $m_i^2$ and $\mu_\phi^2$ are positive quantities.\footnote{The eigenvalues can always be written in this form if at least one of them is negative. In this case, we can choose the minimal one to be the $-\mu_\phi^2$ and express all the others as $m_i^2-\mu_\phi^2$.} The SO(N) symmetry is spontaneously broken due to the negative mass-squared parameters, while the $m_i^2$ are soft-breaking masses. Note that in Ref.\cite{Alanne:2018zjm}, all the $m_i^2$ are assumed to be equal so that the remnant symmetry is exactly SO(N-1). In our setup, we are considering a more general situation in which the soft-breaking terms also break the SO(N-1) symmetry. Using the new basis, the Lagrangian can be rewritten as
\beq\label{SOLag}
\mathcal{L}&=&(D_\mu H)^\dag D^\mu H+\mu^2|H|^2-\lambda|H|^4\nonumber\\
&&+(\partial_\mu \tilde{\Phi})^T \partial^\mu\tilde{\Phi}+\mu_\phi^2(\tilde{\Phi}^T\tilde{\Phi})-\lambda_\Phi(\tilde{\Phi}^T\tilde{\Phi})^2-2\lambda_{H\Phi}|H|^2(\tilde{\Phi}^T\tilde{\Phi})\nonumber\\
&&-\tilde{\Phi}^T\begin{pmatrix}m_1^2&&&&\\&m_2^2&&&\\ &&\ddots&&\\&&&m_{N-1}^2&\\&&&&0\end{pmatrix}\tilde{\Phi}~.
\eeq
The last term of eq.\eqref{SOLag} is a diagonal soft-breaking mass term which gives masses to the Goldstone bosons.
In this case, the $\tilde{\Phi}$ field can be parametrized as
\beq
\tilde{\Phi}=\frac{v_\phi+\phi}{\sqrt{2}}\mathrm{exp}\left(i\frac{\chi^{\hat{a}}T^{\hat{a}}}{v_\phi}\right)\begin{pmatrix}0\\ \vdots\\0\\1\end{pmatrix}=\frac{v_\phi+\phi}{\sqrt{2}}\begin{pmatrix}\frac{\chi^{\hat{a}}}{|\chi^{\hat{a}}|}\sin\left(\frac{|\chi^{\hat{a}}|}{v_\phi}\right)\\ \cos\left(\frac{|\chi^{\hat{a}}|}{v_\phi}\right)\end{pmatrix}
\eeq
where $\hat{a}=1,2,...,N-1$ indicate the broken generators given by
\beq
T^{1}=\begin{pmatrix}0&0&\dots&-i\\ 0&0&\dots&0\\ \vdots&\vdots&\ddots&\vdots\\i&0&\dots&0\end{pmatrix},\quad T^{2}=\begin{pmatrix}0&0&\dots&0\\ 0&0&\dots&-i\\ \vdots&\vdots&\ddots&\vdots\\0&-i&\dots&0\end{pmatrix},\dots,~ T^{N-1}=\begin{pmatrix}0&0&\dots&0\\ \vdots&\vdots&\vdots&\vdots\\ 0&0&\ddots&-i\\ 0&\dots&i&0\end{pmatrix},\nonumber
\eeq
 and $\chi^{\hat{a}}$ are pNGBs which play the role of multi-component DM.
The minimization conditions of the potential are
\beq
\mu_\phi^2=\lambda_\Phi v_\phi^2+\lambda_{H\Phi}v^2,\qquad \mu^2=\lambda v^2+\lambda_{H\Phi}v_\phi^2,
\eeq
and these will determine the VEVs in terms of the model parameters. The mass matrix for $(h,\phi)$ is given by
\beq
M_{h,\phi}^2=\begin{pmatrix}2\lambda v^2&2\lambda_{H\Phi}vv_\phi\\2\lambda_{H\Phi}vv_\phi&2\lambda_\Phi v_\phi^2\end{pmatrix}~.
\eeq
The kinetic term of $\hat{\Phi}$ can be computed by expanding the exponential function and keeping the leading terms up to the quadratic of $\chi^{\hat{a}}$ as follows,
\beq\label{expansion}
\tilde{\Phi}=\frac{v_\phi+\phi}{\sqrt{2}}\begin{pmatrix}\chi^{1}/v_\phi \\ \chi^{2}/v_\phi \\ \vdots\\ \chi^{N-1}/v_\phi \\1-\chi^{\hat{a}}\chi^{\hat{a}}/(2v_\phi^2)\end{pmatrix}+\mathcal{O}(\chi^{\hat{a}}\chi^{\hat{b}}\chi^{\hat{c}})~.
\eeq
Substituting eq.\eqref{expansion} into the kinetic term, we obtain
\beq
\mathcal{L}_{kin}=\frac{1}{2}\partial_\mu\phi \partial^\mu\phi+\frac{(v_\phi+\phi)^2}{2v_\phi^2}\sum_{\hat{a}}(\partial_\mu\chi^{\hat{a}})(\partial^\mu\chi^{\hat{a}})+\dots~.
\eeq
The soft-breaking mass terms can also be computed by the expansion and the result  is
\beq
\mathcal{L}_{soft}\supset-\frac{(v_\phi+\phi)^2}{2v_\phi^2}\sum_{\hat{a}}m_{\hat{a}}^2(\chi^{\hat{a}})^2
\eeq
It is now easy to show that the masses of the pNGBs $\chi^{\hat{a}}$ are $m_{\hat{a}}$, while the trilinear interactions of $\chi^{\hat{a}}$ and $\phi$ are
\beq\label{SON_tril}
\mathcal{L}_{\phi(\chi^{\hat{a}})^2}=\sum_{\hat{a}}\frac{\partial^2\phi}{2v_\phi}(\chi^{\hat{a}})^2-\sum_{\hat{a}}\frac{\phi}{v_\phi}\chi^{\hat{a}}(\partial^2+m_{\hat{a}}^2)\chi^{\hat{a}}.
\eeq
Once we read off the Feynman rules in momentum space, the first term of eq.\eqref{SON_tril} is proportional to the momentum squared of $\phi$, while the second term vanishes when $\chi^{\hat{a}}$ is on-shell.
Since the pNGBs can only communicate with the SM fermions through $\phi$ which is mixing with $h$, the amplitude of the $\chi^{\hat{a}}$-quark scattering vanishes in the zero-momentum transfer limit. Note that when N is an even number, the model is equivalent to the SU(N/2) generalization which has been discussed previously in Ref.\cite{Karamitros:2019ewv}.

We can now try to add some soft-breaking cubic terms to eq.\eqref{SOLag} and see how they affect our results. Without loss of generality, these terms can be written,
\beq\label{cubicterm}
\mathcal{L}_{\Phi^3}=-\kappa_{ijk}\tilde{\Phi}^i\tilde{\Phi}^j\tilde{\Phi}^k~,
\eeq
 where $i,j,k=1,2,...,N$ represent the indices of the fundamental representation, while $\kappa_{ijk}$ are real parameters with 1 unit of mass dimension. Using the expansion given by eq.\eqref{expansion}, we can analyze these terms separately for different components, as follows
 \begin{itemize}
 	\item The $\kappa_{\hat{a}\hat{b}\hat{c}}\tilde{\Phi}^{\hat{a}}\tilde{\Phi}^{\hat{b}}\tilde{\Phi}^{\hat{c}}\sim\frac{(v_\phi+\phi)^3}{2\sqrt{2}v_\phi^3}\kappa_{\hat{a}\hat{b}\hat{c}}\chi^{\hat{a}}\chi^{\hat{b}}\chi^{\hat{c}}$ terms include trilinear interactions among the pNGBs. They can be categorized to the following three types:
 	\begin{enumerate}
 	   \item If $\hat{a}=\hat{b}=\hat{c}$, then $\chi^{\hat{a}}$ cannot be a DM candidate since this interaction leads to DM decay.
 	
 	   \item If $\hat{a}=\hat{b}\neq\hat{c}$, $\chi^{\hat{a}}$ is a viable DM candidate, since it is protected by $Z_2$ parity $\chi^{\hat{a}}\to -\chi^{\hat{a}}$, while $\chi^{\hat{c}}$ is unstable and thus can not be DM. In this case, the $\chi^{\hat{a}}$ DM still enjoys the cancellation mechanism since $\chi^{\hat{c}}$ does not mix with the radial mode or the Higgs boson $h$ to leading order. \footnote{Note that $\chi^{\hat{c}}$ does mix with the radial mode at the 1-loop level. Therefore, it can give rise to a non-vanishing $\chi^{\hat{a}}$-quark scattering amplitude even in the $t=0$ limit. However, such an effect can be naturally small since it is suppressed by a loop factor.}
 	
 	   \item If $\hat{a}\neq\hat{b}\neq\hat{c}$, all $\chi$ fields are stable DM candidates, unless one of them has a mass larger than the sum of the other two, in which case it will decay into the lighter states. The stability of DM can also be guaranteed by introducing a $Z_2\times Z_2$ symmetry. We can let $\chi^{\hat{a}},~\chi^{\hat{b}}$ being odd under the first $Z_2$, while let $\chi^{\hat{b}},~\chi^{\hat{c}}$ being odd under the second $Z_2$.
    \end{enumerate}
 \item The $\kappa_{\hat{a}\hat{b}N}\tilde{\Phi}^{\hat{a}}\tilde{\Phi}^{\hat{b}}\tilde{\Phi}^{N}\sim\frac{(v_\phi+\phi)^3}{2\sqrt{2}v_\phi^2}\kappa_{\hat{a}\hat{b}N}\chi^{\hat{a}}\chi^{\hat{b}}[1-\sum_{\hat{c}}\chi^{\hat{c}}\chi^{\hat{c}}/(2v_\phi^2)+\cdots]$ terms lead to mass mixing between $\chi^{\hat{a}}$ and $\chi^{\hat{b}}$, and in addition their trilinear interactions with the radial mode. When we add only these terms to the model, they definitely violate the cancellation mechanism for $\chi^{\hat{a}}$ and $\chi^{\hat{b}}$, since the trilinear coupling from the potential term can no longer cancel the contribution from the kinetic term. Moreover, once the full mass matrix of $\chi^{\hat{a}}$ and $\chi^{\hat{b}}$ has a negative eigenvalue, the vacuum we have chosen is not stable anymore.
 \item The $\kappa_{\hat{a}NN}\tilde{\Phi}^{\hat{a}}(\tilde{\Phi}^{N})^2\sim\frac{(v_\phi+\phi)^3}{2\sqrt{2}v_\phi}\kappa_{\hat{a}NN}\chi^{\hat{a}}[1-\sum_{\hat{c}}\chi^{\hat{c}}\chi^{\hat{c}}/(2v_\phi^2)+\cdots]^2$ terms contain tadpoles of $\chi^{\hat{a}}$, which means we have chosen a wrong vacuum configuration.
 In this case, a more complicated formulation of the vacuum is required, which is beyond the scope of this work but worth studying in the future.
 \item The $\kappa_{NNN}(\tilde{\Phi}^{N})^3\sim\frac{(v_\phi+\phi)^3}{2\sqrt{2}}\kappa_{NNN}(1-\chi^{\hat{c}}\chi^{\hat{c}}/(2v_\phi^2))^3$ terms generate extra mass terms and trilinear interactions for each pNGB. If these terms are added individually, the cancellation mechanism is violated for all pNGBs.
 \item A special combination  $3\kappa_{\hat{a}\hat{a}N}\tilde{\Phi}^{\hat{a}}\tilde{\Phi}^{\hat{a}}\tilde{\Phi}^{N}+\kappa_{NNN}(\tilde{\Phi}^{N})^3$, with $\kappa_{\hat{a}\hat{a}N}=\frac{1}{2}\kappa_{NNN}$, cancels the extra quadratic of $\chi^{\hat{a}}$ and, therefore, the cancellation still works for $\chi^{\hat{a}}$.
 \end{itemize}

On the other hand, the symmetric tensor $\kappa_{ijk}$ can be separated into two parts which might originates from different sources. One is a spurion $\kappa_i$ in the fundamental representation, so that $\kappa_{ijk}$  can be written as $\kappa_{ijk}=\kappa_i\delta_{jk}+\kappa_j\delta_{ik}+\kappa_k\delta_{ij}$. The other source of $\kappa_{ijk}$ is a spurion in a symmetric three-index irreducible representation of SO(N), which is denoted as $\tilde{\kappa}_{ijk}$, with conditions $\tilde{\kappa}_{ijk}\delta^{ij}=\tilde{\kappa}_{ijk}\delta^{jk}=\tilde{\kappa}_{ijk}\delta^{ik}=0$. In principle, we should not expect the coefficients of operators originating from these two different sources to be related. Note that, in the first case, only the $\kappa_N$ component is permitted to be non-zero, otherwise the vacuum configuration is incorrectly chosen. However, the $\kappa_N$ component leads to a violation of the cancellation mechanism for all the pNGBs, so this case will not be considered in this work.

In the original U(1) (or SO(2)) model, only a single pNGB appears so no $\kappa_{\hat{a}\hat{b}\hat{c}}\tilde{\Phi}^{\hat{a}}\tilde{\Phi}^{\hat{b}}\tilde{\Phi}^{\hat{c}}$ term can be added without violating the cancellation. Although it can include a term such as $\frac{3}{2}(\tilde{\Phi}^{1})^2\tilde{\Phi}^{2}+(\tilde{\Phi}^{2})^3$, which preserves the cancellation,\footnote{In the formulation of U(1) symmetry with the complex scalar $S$, this term is just the case that letting $\kappa_1=\kappa_2+9\kappa_3=0$ in eq.\eqref{U1cubicsoft}.} as we have mentioned, it requires an unnatural combination of the two unrelated terms from different sources. The simplest model allowing $\kappa_{\hat{a}\hat{b}\hat{c}}\tilde{\Phi}^{\hat{a}}\tilde{\Phi}^{\hat{b}}\tilde{\Phi}^{\hat{c}}$ terms is the SO(3) model which contains two pNGBs. To find out the most general cubic terms naturally preserving the cancellation, we consider the scenario in which $\kappa_{ijk}$ stems from a spurion in an irreducible symmetric three-index representation of SO(3). The resulting cubic terms can be parametrized as
\beq\label{SO3cubic}
-\mathcal{L}_{\Phi^3}&=&+\tilde{\kappa}_{111}(\tilde{\Phi}^{1})^3+3\tilde{\kappa}_{122}\tilde{\Phi}^{1}(\tilde{\Phi}^{2})^2-3(\tilde{\kappa}_{111}+\tilde{\kappa}_{122})\tilde{\Phi}^{1}(\tilde{\Phi}^{3})^2\nonumber\\
&&+3\tilde{\kappa}_{112}(\tilde{\Phi}^{1})^2\tilde{\Phi}^{2}+\tilde{\kappa}_{222}(\tilde{\Phi}^{2})^3-3(\tilde{\kappa}_{112}+\tilde{\kappa}_{222})\tilde{\Phi}^{2}(\tilde{\Phi}^{3})^2\nonumber\\
&&+3\tilde{\kappa}_{113}(\tilde{\Phi}^{1})^2\tilde{\Phi}^{3}+3\tilde{\kappa}_{223}(\tilde{\Phi}^{2})^2\tilde{\Phi}^{3}-(\tilde{\kappa}_{113}+\tilde{\kappa}_{223})(\tilde{\Phi}^{3})^3+6\tilde{\kappa}_{123}\tilde{\Phi}^{1}\tilde{\Phi}^{2}\tilde{\Phi}^{3}~.
\eeq
 Note that independent degrees of freedom of $\tilde{\kappa}_{ijk}$ is seven, which match the number of dimensions for the symmetric three-index representation. We assume that the chosen vacuum is stable, so that following conditions should be satisfied
 \beq\label{vacstab}
 \tilde{\kappa}_{111}+\tilde{\kappa}_{122}=\tilde{\kappa}_{112}+\tilde{\kappa}_{222}=0~,
 \eeq
 otherwise tadpole terms of pNGBs will be generated. If we require that at least one pNGB particle, $\chi^1$, is a stable DM candidate, following conditions should also be satisfied
 \beq\label{chi1stable}
 \tilde{\kappa}_{111}= \tilde{\kappa}_{122}= \tilde{\kappa}_{123}=0~.
 \eeq
 Finally, if we want the cancellation mechanism to work for $\chi^1$, we requires
 \beq\label{cancond113223}
  \tilde{\kappa}_{113}=-\frac{1}{3} \tilde{\kappa}_{223}~,
 \eeq
 Substituting these conditions into eq.\eqref{SO3cubic} and expanding in series of $\chi^{\hat{a}}$, we obtain the explicit expression for the cubic terms,
 \beq
-\mathcal{L}_{\Phi^3}&=&\tilde{\kappa}_{112}[3\tilde{\Phi}^{2}(\tilde{\Phi}^{1})^2-(\tilde{\Phi}^{2})^3]+\tilde{\kappa}_{223}\left[-\tilde{\Phi}^{3}(\tilde{\Phi}^{1})^2+3\tilde{\Phi}^{3}(\tilde{\Phi}^{2})^2-\frac{2}{3}(\tilde{\Phi}^{3})^3\right]\nonumber\\
&=&\frac{(v_\phi+\phi)^3}{2\sqrt{2}v_\phi^3}\left\{\tilde{\kappa}_{112}[3(\chi^1)^2\chi^2-(\chi^2)^3]+\tilde{\kappa}_{223}\left[-\frac{2}{3}v_\phi^3+4v_\phi(\chi^2)^2\right]+\dots\right\}~.
 \eeq
 Note that the condition eq.\eqref{chi1stable} can be naturally satisfied if we assume a $Z_2$ symmetry under which $\tilde{\Phi}^1$ is odd. On the other hand, the condition eq.\eqref{cancond113223} is not automatically satisfied in the most general case. However, a special case that $\tilde{\kappa}_{113}=\tilde{\kappa}_{223}=0$, can be naturally satisfied by assuming a $Z_2$ symmetry under which $\tilde{\Phi}^3$ is odd. In this case, the $Z_2$ symmetry is spontaneously broken by the VEV of $\tilde{\Phi}^3$, so that eq.\eqref{cancond113223} can still be slightly violated at loop-level. We expect that the loop-level violation is small and does not lead to significant effect in the direct detection process. If we further assume that $\tilde{\kappa}_{112}= \tilde{\kappa}_{222}=0$, then $\chi^2$ can also be a stable DM particle, but it does not preserve the cancellation unless $\tilde{\kappa}_{113}$ and $\tilde{\kappa}_{223}$ also vanish. In conclusion, the SO(3) model is the minimal model which can include some soft-breaking cubic terms without violating the cancellation mechanism for the DM candidates

Finally we want to make some comments on the second condition in eq.\eqref{vacstab}, which is $\tilde{\kappa}_{112}+\tilde{\kappa}_{222}=0$. As a low energy effective theory, we just assume this condition by hand, for the consistency of the chosen vacuum. However, a non-vanishing $\tilde{\kappa}_{112}$ which respects this condition can also be automatically generated if we consider a UV completion of the spurion $\tilde{\kappa}_{ijk}$. We assume that all the soft-breaking terms coming from a single real scalar field, $K_{ijk}$, which is in an irreducible symmetric 3-index representation of the SO(3) symmetry. The $K_{ijk}$ field can couple to the $\Phi^i$ field through following renormalizable potential terms,
\beq\label{VKPhi}
V_{K\Phi}&=&\lambda_{K(\Phi)^3}K_{ijk}\Phi^i\Phi^j\Phi^k+\lambda_{(K)^2(\Phi)^2}K_{ijk}K_l^{jk}\Phi^i\Phi^l+\lambda'_{(K)^2(\Phi)^2}\varepsilon_{imn}\varepsilon_{jkl}K^{mkp}K^{nl}_p \Phi^i\Phi^j\nonumber\\
&&+\lambda_{(K)^3\Phi}K_{ijk}K^{jlm}K^k_{lm}\Phi^i+\lambda'_{(K)^3\Phi}\varepsilon^{jlp}\varepsilon^{kmq}K_{ijk}K_{lmn}K_{pq}^n\Phi^i\nonumber\\
&&+\lambda''_{(K)^3\Phi}\varepsilon_{ijk}\varepsilon_{lmq}K^{jln}K^{kmp}K^q_{np}\Phi^i~,
\eeq
where $\varepsilon_{ijk}$ is the Levi-Civita symbol. When $K_{ijk}$ acquires a non-trivial VEV, $\langle K_{ijk}\rangle\equiv \tilde{\kappa}_{ijk}/\lambda_{K(\Phi)^3}$, the first term in eq.\eqref{VKPhi} generates the cubic term given in eq.\eqref{cubicterm}. The second and third terms generate soft-breaking masses $\Delta M_{ij}^2$, while the terms in the second and third lines generate soft-breaking linear terms for $\Phi^i$. If only $K_{112}$ and $K_{222}$ have non-zero VEV, $\langle K_{112}\rangle=-\langle K_{222}\rangle\equiv\kappa/\lambda_{K(\Phi)^3}$ (see Appendix \ref{app2} for more details about the vacuum configuration), we can check that all the linear soft-breaking terms vanish, while the soft-breaking mass matrix of $\Phi^i$ is
\beq
\Delta M_{ij}^2=\begin{pmatrix}\frac{2\lambda_{(K)^2(\Phi)^2}}{(\lambda_{K(\Phi)^3})^2}\kappa^2&&\\&\frac{2\lambda_{(K)^2(\Phi)^2}}{(\lambda_{K(\Phi)^3})^2}\kappa^2&\\&&-\frac{2\lambda'_{(K)^2(\Phi)^2}}{(\lambda_{K(\Phi)^3})^2}\kappa^2\end{pmatrix}~,
\eeq
which is consistent with the structure of eq.\eqref{diagmass}. Therefore, we have found a UV completion of the soft-breaking cubic term for the SO(3) model, in which a pNGB DM candidate can preserve the cancellation property.

In the case that $ \tilde{\kappa}_{113}\propto\langle K_{113}\rangle\neq0$, the linear term induced by the $KKK\Phi$ coupling is usually non-vanishing, so that our previous discussion which only included the cubic term of $\Phi^i$ was incomplete. Since the situation is much more complicated, we leave it for future research.

\subsection{The second proof}
In our second proof of the cancellation mechanism, we use the linear representation eq.\eqref{linear}. The mass term and trilinear couplings from the potential are given by
\beq
-\mathcal{L}&\supset&\lambda_Hv^2h^2+2\lambda_{SH}vv_shs+\lambda_Sv_s^2s^2+\frac{1}{2}m_\chi^2\chi^2\nonumber\\
&&+\lambda_{SH}vh\chi^2+\lambda_{S}v_ss\chi^2~,
\eeq
which can be written in the following quadratic form
\beq\label{lag3}
-\mathcal{L}&\supset&\left(\lambda_H-\frac{\lambda_{SH}^2}{\lambda_S}\right)v^2h^2+\frac{1}{\lambda_S}(\lambda_{SH}vh+\lambda_Sv_ss)^2+\frac{1}{2}m_\chi^2\chi^2\nonumber\\
&&+(\lambda_{SH}vh+\lambda_Sv_s s)\chi^2
\eeq
We then find that the combination of $h$ and $s$ that couples to the $\chi$ appears as a quadratic form in the potential. If we define a new scalar, $\phi\equiv (\lambda_{SH}vh+\lambda_Sv_ss)/\lambda_Sv_s$, and rewrite the Lagrangian in terms of $h$ and $\phi$ as
\beq
\mathcal{L}&\supset&\frac{1}{2}\left[1+\left(\frac{\lambda_{SH}v}{\lambda_Sv_s}\right)^2\right]\partial_\mu h\partial^\mu h+\frac{1}{2}\partial_\mu\phi\partial^\mu\phi-\frac{\lambda_{SH}v}{\lambda_Sv_s}\partial_\mu h\partial^\mu\phi\nonumber\\
&&-\left(\lambda_H-\frac{\lambda_{SH}^2}{\lambda_S}\right)v^2h^2-\lambda_Sv_s^2\phi^2-\frac{1}{2}m_\chi^2\chi^2-\lambda_Sv_s\phi \chi^2,
\eeq
we see that the mass terms are already diagonalized in this form. On the other hand, the kinetic terms are not canonical anymore and there is also a kinetic mixing term between $h$ and $\phi$ generated in this basis.
We can treat the kinetic mixing term as an interacting vertex which is endowed with a value
\beq
ig_{h\phi}q^2=-i\frac{\lambda_{SH}v}{\lambda_Sv_s}q^2
\eeq
where $q^\mu$ is the momentum of $\phi$. The propagators of $h$ and $\phi$ can be read off from their non-canonical kinetic and mass terms as
\beq
D_\phi(q)=\frac{i}{q^2-m_\phi^2},\qquad D_h(q)=\frac{i}{\xi_hq^2-m_h^2},
\eeq
where $\xi_h=1+(\lambda_{SH}v/\lambda_Sv_s)^2$, $m_\phi^2=2\lambda_Sv_s^2$ and $m_h^2=2(\lambda_H-\lambda_{SH}^2/\lambda_S)v^2$. Note that in this form the SM fermions only couple to the $h$ field, while the DM $\chi$ only couples to the $\phi$ field. The only portal connecting these two sectors is the kinetic mixing vertex whose strength is proportional to the momentum squared of the mediator, which is just the $t$ variable (see FIG.\ref{fig1c}). Therefore we see that the amplitude of the $\chi$-quark scattering,
\beq
i\mathcal{M}=(2i\lambda_Sv_s)\frac{i}{t-m_\phi^2}\left(-i\frac{\lambda_{SH}v}{\lambda_Sv_s}t\right)\frac{i}{\xi_ht-m_h^2}\left(-i\frac{m_q}{v}\right)\bar{u}(k_2)u(p_2)+...
\eeq
vanishes in the $t\to0$ limit.

The lesson we can learn from the second proof is that the cancellation mechanism relies on the special structures in the masses and interactions of the scalar mediators. The condition of the cancellation is that the combination of the mediators that appears in the trilinear coupling with the DM has no mass mixing with the SM Higgs boson $h$. This inspires us to look for the same structure in the vector-portal models. When the gauge symmetry is spontaneously broken by the Higgs mechanism, the masses of the gauge boson are generated by the gauge interaction of the Higgs field. If the representation of gauge symmetries for the DM field is the same as the a new Higgs field, which breaks some new gauge groups, then DM field will couple to the new gauge bosons with the same combination of mediators as the new Higgs field. Therefore, the cancellation in the vector-portal models can be achieved by the same method as in the Higgs-portal models. In the following section, we give two examples of the cancellation mechanism in vector-portal models in details.

\section{Cancellation mechanism in the vector-portal models}\label{sect.vec}

\subsection{The $\textrm{SU(2)}_L\times \textrm{U(1)}_Y\times \textrm{U(1)}_X$ model}
Firstly, we consider a simple extension of the SM with a gauged U(1)$_X$ symmetry as a toy model for illustrating the vector-portal cancellation mechanism. In our setup, besides the SM Higgs doublet $H$, we introduce a new Higgs field $\Phi$ and a Dirac fermion DM field, $\Psi$ which are both in the representation $(2,~1/2,~1)$ of the gauge symmetry $\textrm{SU(2)}_L\times \textrm{U(1)}_Y\times \textrm{U(1)}_X$. The covariant derivatives of $H,~\Phi,$ and $\Psi$ are given by
\beq
D_\mu H&=&\left[\partial_\mu-i\begin{pmatrix}eA_\mu+\frac{gc_{2W}}{2c_W}Z_\mu&\frac{gW^+_\mu}{\sqrt{2}}\\ \frac{gW^-_\mu}{\sqrt{2}}&\frac{g}{2c_W}Z_\mu\end{pmatrix}\right]\begin{pmatrix}H^+\\H^0\end{pmatrix},\\
D_\mu \Phi&=&\left[\partial_\mu-i\begin{pmatrix}eA_\mu+\frac{gc_{2W}}{2c_W}Z_\mu+g_XX_\mu&\frac{gW^+_\mu}{\sqrt{2}}\\ \frac{gW^-_\mu}{\sqrt{2}}&\frac{g}{2c_W}Z_\mu+g_XX_\mu\end{pmatrix}\right]\begin{pmatrix}\Phi^+\\
	 \Phi^0\end{pmatrix},\\
 D_\mu \Psi&=&\left[\partial_\mu-i\begin{pmatrix}eA_\mu+\frac{gc_{2W}}{2c_W}Z_\mu+g_XX_\mu&\frac{gW^+_\mu}{\sqrt{2}}\\ \frac{gW^-_\mu}{\sqrt{2}}&\frac{g}{2c_W}Z_\mu+g_XX_\mu\end{pmatrix}\right]\begin{pmatrix}\chi^+\\ \chi\end{pmatrix},
\eeq
where $g_X$ is the gauge coupling of U(1)$_X$. The $\chi$ field, which is the neutral component of $\Psi$, is the DM candidate. After $H$ and $\Phi$ acquire their VEVs, defined as
\beq
\langle H\rangle=\begin{pmatrix}0\\ \frac{v_h}{\sqrt{2}}\end{pmatrix}~,\qquad\langle \Phi\rangle=\begin{pmatrix}0\\ \frac{v_\phi}{\sqrt{2}}\end{pmatrix}~,
\eeq
the $\textrm{SU(2)}_L\times \textrm{U(1)}_Y\times \textrm{U(1)}_X$ gauge symmetries are spontaneously broken and both the $Z_\mu$ and $X_\mu$ boson become massive. The masses terms of the neutral gauge bosons are
\beq
-\mathcal{L}_m=\frac{1}{2}\left(\frac{gv_h}{2c_W}Z_\mu\right)^2+\frac{1}{2}\left(\frac{gv_\phi}{2c_W}Z_\mu+g_Xv_\phi X_\mu\right)^2.
\eeq
The mass matrix of the gauge field $(Z_\mu,~X_\mu)$ can be read off as follows,
\beq
M_{g}^2=\begin{pmatrix}\frac{g^2(v_h^2+v_\phi^2)}{4c_W^2}&\frac{gg_Xv_\phi^2}{2c_W}\\ \frac{gg_Xv_\phi^2}{2c_W}&g_X^2v_\phi^2\end{pmatrix}~.
\eeq
It can be diagonalized by an orthogonal transformation $U$ as $M_{diag}^2=UM_{g}^2U^T$, while the mass eigenstate is $(\hat{Z}_\mu,~\hat{Z}^\prime_\mu)=(Z_\mu,~X_\mu)U^T$. The mass of the $W^\pm_\mu$ boson is just
\beq
m_W=\frac{g\sqrt{v_h^2+v_\phi^2}}{2},
\eeq
and we can define $v=\sqrt{v_h^2+v_\phi^2}\approx246.22$~GeV as the SM VEV of the Higgs field. It constrains the values of $v_h,v_\phi$ to be $v_h,v_\phi\leq v$.
\begin{figure}[tb]
	\begin{subfigure}[b]{.3\textwidth}
		\centering
		\includegraphics[width=\textwidth]{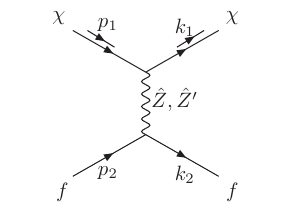}
		\caption[]{}\label{fig2a}
	\end{subfigure}
	\begin{subfigure}[b]{.3\textwidth}
		\centering
		\includegraphics[width=\textwidth]{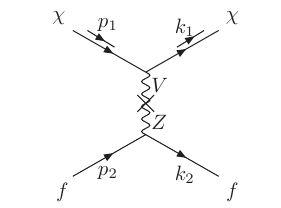}
		\caption{}\label{fig2b}
	\end{subfigure}
	\caption{Plots (a) and (b) represent two different points of view of the DM-$f$ (SM fermion) scattering through the vector-portal. In Plot (a), the mediators $\hat{Z}$ and $\hat{Z}'$ are mass eigenstats, and their amplitudes cancel each others miraculously. In Plot (b), there is only one diagram whose mediator involves a mixing between the $V_\mu$ and $Z_\mu$ fields. }
	\label{feyndiag2}
\end{figure}
The gauge interactions of the DM candidate $\chi$ are given by
\beq
\mathcal{L}_g=\left(\frac{g}{2c_W}Z_\mu+g_XX_\mu\right)\overline{\chi}\gamma^\mu\chi=\begin{pmatrix}\frac{g}{2c_W}~,&g_X\end{pmatrix}U^T\begin{pmatrix}\hat{Z}_\mu\\ \hat{Z}^\prime_\mu\end{pmatrix}\overline{\chi}\gamma^\mu\chi
\eeq
The diagram of $\chi$-$f$ (SM fermion) scattering is shown in FIG.\ref{fig2a}~, and its corresponding amplitude can be computed as
\beq
i\mathcal{M}&=&\bar{u}_\chi(k_1)\gamma^\mu u_\chi(p_1)i\begin{pmatrix}\frac{g}{2c_W}~,&g_X\end{pmatrix}U^T(-i)\begin{pmatrix}\frac{g_{\mu\nu}-\frac{q^\mu q^\nu}{m_Z^2}}{t-m_{Z}^2}&0\\ 0&\frac{g_{\mu\nu}-\frac{q^\mu q^\nu}{m_{Z'}^2}}{t-m_{Z'}^2}\end{pmatrix} U \begin{pmatrix}1\\0\end{pmatrix}\nonumber\\
&&\times i\bar{u}_f(k_2)iG_{Z\bar{f}f}^{\nu} u_f(p_2)~,
\eeq
where $G_{Z\bar{f}f}^{\nu}=g_{Z\bar{f}f}^V\gamma^\nu-g_{Z\bar{f}f}^A\gamma^\nu\gamma^5$.
In the limit $q^\mu=(\frac{\vec{p}_1^2-\vec{k}_1^2}{2m_{DM}}+...,\vec{p}_1-\vec{k}_1)\to0$ and $t=q^2\to0$, the amplitude becomes
\beq
i\mathcal{M}&\to&-i\bar{u}_\chi(k_1)\gamma_\mu u_\chi(p_1)\begin{pmatrix}\frac{g}{2c_W}~,&g_X\end{pmatrix}U^T\begin{pmatrix}\frac{1}{m_{Z}^2}&0\\ 0&\frac{1}{m_{Z'}^2}\end{pmatrix} U\begin{pmatrix}1\\0\end{pmatrix}\bar{u}_f(k_2)G_{Z\bar{f}f}^{\mu} u_f(p_2)\nonumber\\
&=&-i\bar{u}_\chi(k_1)\gamma_\mu u_\chi(p_1)\bar{u}_f(k_2)G_{Z\bar{f}f}^{\mu} u_f(p_2)\begin{pmatrix}\frac{g}{2c_W}~,&g_X\end{pmatrix}(M_g^2)^{-1}\begin{pmatrix}1\\0\end{pmatrix}\nonumber\\
&=&-\left(\frac{i}{\textrm{det}(M_g^2)}\right)\bar{u}_\chi(k_1)\gamma_\mu u_\chi(p_1)\bar{u}_f(k_2)G_{Z\bar{f}f}^{\mu}u_f(p_2)\nonumber\\
&&\times\begin{pmatrix}\frac{g}{2c_W}~,&g_X\end{pmatrix}\begin{pmatrix}g_X^2v_\phi^2&-\frac{gg_Xv_\phi^2}{2c_W}\\ -\frac{gg_Xv_\phi^2}{2c_W}&\frac{g^2(v_h^2+v_\phi^2)}{4c_W^2}\end{pmatrix}\begin{pmatrix}1\\0\end{pmatrix}=0~,
\eeq
which implies a suppressed direct detection signal.
The cancellation mechanism in this case can also be proved by the second method we used for the Higgs-portal model. Note that the structures of the gauge interaction and the mass terms are similar to the Higgs-portal model. We can define a vector $V_\mu=(gZ_\mu/(2c_W)+g_XX_\mu)/g_X$, and rewrite the Lagrangian in terms of $Z_\mu$ and $V_\mu$ as
\beq\label{lag_ZV}
\mathcal{L}&\supset&-\frac{1}{4}\left(1+\frac{g^2}{4c_W^2g_X^2}\right)(\partial_\mu Z_\nu-\partial_\nu Z_\mu)^2-\frac{1}{4}(\partial_\mu V_\nu-\partial_\nu V_\mu)^2\nonumber\\
&&+\frac{g}{4c_Wg_X}(\partial_\mu V_\nu-\partial_\nu V_\mu)(\partial^\mu Z^\nu-\partial^\nu Z^\mu)\nonumber\\
&&+\sum_f\bar{f}(g_{Z\bar{f}f}^V\gamma^\mu+g_{Z\bar{f}f}^A\gamma^\mu\gamma^5)fZ_\mu+g_X\overline{\chi}\gamma^\mu\chi V_\mu~.
\eeq
The second line of eq.\eqref{lag_ZV} is a kinetic mixing term for $V_\mu$ and $Z_\mu$, which can be treated as an interacting vertex in terms of the momentum of $V_\mu$. Since the standard model fermions do not couple to the $V_\mu$ field, while the dark matter $\chi$ only couples to $V_\mu$, their scattering can only be induced by the kinetic mixing vertex (see the diagram shown in FIG.\ref{fig2b}), which vanishes in the zero-momentum transfer limit.\footnote{In the $(Z_\mu,~V_\mu)$ basis, the form of eq.\eqref{lag_ZV} matches the structure of eq.(B1) in Appendix B of Ref.\cite{Liu:2017lpo}, which have shown a proof for the cancellation.}

\begin{figure}[tb]
	\centering
	\includegraphics[width=0.3\textwidth]{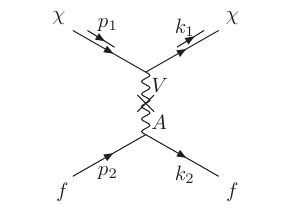}
	\caption{The leading diagram of the $\chi-f$ scattering when there is a generic kinetic mixing term for the gauge fields.}
	\label{feyndiag3}
\end{figure}
\begin{figure}[tb]
	\centering
	\includegraphics[width=0.7\textwidth]{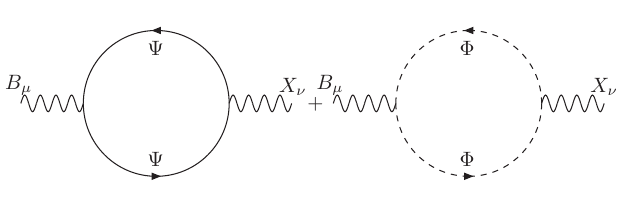}
	\caption{The 1-loop corrections of the $B_{\mu\nu}X^{\mu\nu}$ terms.}
	\label{feyndiag4}
\end{figure}

So far, the cancellation mechanism seems to work well for the vector-portal model. However, the cancellation is violated if the model includes a generic kinetic mixing between the $B_\mu$ and the $X_\mu$ fields in the original Lagrangian, i.e.,
\beq
\mathcal{L}\supset-\frac{1}{4}B_{\mu\nu}B^{\mu\nu}-\frac{1}{4}X_{\mu\nu}X^{\mu\nu}-\frac{s_\epsilon}{2}B_{\mu\nu}X^{\mu\nu}~,
\eeq
where $s_\epsilon\equiv \sin\epsilon$ is a parameter characterizing the kinetic mixing. When we rewrite the Lagrangian using $V_\mu$, this term leads to a kinetic mixing between the $V_\mu$ and the photon field $A_\mu$:
\beq
\mathcal{L}\supset-\frac{s_\epsilon c_W}{2}A_{\mu\nu}V^{\mu\nu}~,
\eeq
which can induce a new diagram in the $\chi$-quark scattering mediated by the photon. See the diagram shown in Fig.\ref{feyndiag3}. The amplitude can be computed as
\beq
i\mathcal{M}&\sim& ig_X\bar{u}_\Psi(k_1)\gamma^\mu u_\Psi(p_1)\frac{-i(g_{\mu\rho}-\frac{q_\mu q_\rho}{m_V^2})}{t-m_V^2}[-is_\epsilon c_W(tg^{\rho\sigma}-q^\rho q^\sigma)]\frac{-ig_{\sigma\nu}}{t}\nonumber\\
&&(iQ_fe)\times\bar{u}_f(k_2)\gamma^\nu u_f(p_2).
\eeq
Since the photon is massless, it has a pole at $q^2=t=0$ and therefore leads to a non-vanishing amplitude when taking $q^\mu\to0$. Although the amplitude is suppressed by the small kinetic mixing parameter $s_\epsilon$, it is still too large to accommodate the direct detection measurement since the process is mediated by the vector boson $V_\mu$ whose mass is smaller than the weak scale. The lightness of $V_\mu$ is due to the fact that $gg_Xv_\phi^2$ should be much smaller than $g^2v^2$ otherwise the $\rho$ parameter will significantly deviate from $1$. We use the fact that $v_\phi\leq v$ and assume that $g_X$ is not too large, so that $m_V\sim g_Xv_\phi$ should be smaller than $m_Z$. The $\chi$-proton cross section corresponding to Fig.\ref{feyndiag3} process is
\beq\label{DDXsect}
\sigma_{V,\chi p}\approx \frac{m_p^2m_\chi^2}{\pi(m_\chi+m_p)^2}\left(\frac{ec_Wg_Xs_\epsilon}{m_V^2}\right)^2\approx8\times 10^{-38}~\textrm{cm}^2\times\left(\frac{100~\textrm{GeV}}{v_\phi}\right)^4\left(\frac{s_\epsilon}{g_X}\right)^2~.
\eeq
Comparing to the current direct detection bound, $\sigma_{\chi N}\sim 4\times 10^{-46}~\textrm{cm}^2$, for $m_\chi\sim1$~TeV~\cite{PandaX-4T:2021bab}, the kinetic mixing must be fine-tuned such that $(s_\epsilon/g_X)\lesssim10^{-4}$, in order to avoid the stringent direct detection bound. It is unnatural to have such a small kinetic mixing because, in principle, this term can be generated by $\Psi$ and $\Phi$ loops (see Fig.\ref{feyndiag4}). It is easy to estimate the 1-loop correction of the kinetic mixing as
\beq
\delta s_\epsilon(\mu)\sim \frac{g'g_X}{48\pi^2}\left[4\ln\left(\frac{\mu^2}{m_\Psi^2}\right)-\ln\left(\frac{\mu^2}{m_\Phi^2}\right)\right].
\eeq
If we consider that in some higher energy scale $\mu\sim \Lambda_{UV}$, there is a UV completion of the model such that $s_\epsilon$ vanishes, then in the low energy regime the kinetic mixing parameter is generated as
\beq
\frac{\delta s_\epsilon^{IR}}{g_X}\sim -0.004\times\ln\left(\frac{\Lambda_{UV}}{m_\Psi}\right)~.
\eeq
If the logarithm is of order unity, the order of magnitude for $|s_\epsilon/g_X|$ is $10^{-3}\sim10^{-2}$, which is much larger than its upper bound implied by the direct detection data. On the other hand, if there is a generic kinetic mixing parameter, $s_{\epsilon}^{(0)}$, and we require a $(s_\epsilon^{IR}/g_X)\lesssim10^{-4}$ in low energy scale, $(s_{\epsilon}^{(0)}/g_X)$ should be fine-tuned to cancel out the loops-induced contribution, which is at least an order of magnitude larger than $(s_\epsilon^{IR}/g_X)$. In conclusion, if we want this model to avoid the stringent direct detection constraint, we need to tolerate some fine-tunings of the kinetic mixing parameter.

\subsection{The $\textrm{SU(2)}_L\times \textrm{U(1)}_Y\times \textrm{U(1)}_{B-L}\times \textrm{U(1)}_X$ model}
If we look at the argument for why the cross section suppression fails in the previous model, the main subtlety is that the VEV $v_\phi$ is constrained to be small since it serves as a part of the $\textrm{SU}(2)_L\times \textrm{U}(1)_Y$ symmetry breaking. This inspires us to find other vector-portal models that suffer from stringent direct detection bound, and to consider whether the cancellation mechanism is able to work in these cases. A potential candidate is the $\textrm{U(1)}_{B-L}$ extension of SM~\cite{Marshak:1979fm,Mohapatra:1980qe,Wetterich:1981bx,Masiero:1982fi,Buchmuller:1991ce}. In this case, all SM fermions are assigned $\textrm{U(1)}_{B-L}$ charges equal to their baryon or lepton numbers. Three right-handed neutrinos are added to ensure the gauge symmetries are free from anomalies, which as bonus, generate the neutrino masses at the same time. We also introduce a Dirac fermionic DM, $\chi$, which is charged under the $\textrm{U(1)}_{B-L}$. The gauge boson of the $\textrm{U(1)}_{B-L}$ model, $Z'_\mu$,  can be a mediator generating the DM-quark scattering processes. Since these processes are vector mediated, the cross sections will be large and, thus, direct detection will place a stringent bound on the mass scale of $Z'_\mu$. Using the results from Ref.\cite{Duerr:2015wfa}, the DM-nucleon scattering cross section, mediated by $Z'_\mu$, is
\beq
\sigma_{\chi N}^{\mathrm{SI}}\approx1.2 \times 10^{-40}\left(\frac{1~\mathrm{TeV}}{m_{Z'}/(\sqrt{n}g_{B-L})}\right)^{4} \mathrm{~cm}^{2}~,
\eeq
where $n$ is the B-L charge of $\chi$. For a DM mass of around $1$~TeV, current bounds from XENON1T~\cite{Aprile:2018dbl}, LUX~\cite{Akerib:2016vxi}, and PandaX-4T~\cite{PandaX-4T:2021bab} are $\sigma_{\chi N}^{\mathrm{SI}}\lesssim 10^{-45}\mathrm{~cm}^{2}$, which implies $m_{Z'}/(\sqrt{n}g_{B-L})\gtrsim20$~TeV. For a comparison, the LEP bound on $m_{Z'}/g_{B-L}$ is about $6.9$~TeV~\cite{Schael:2013ita,Heeck:2014zfa,Okada:2018ktp}, while the bound from LHC run-2 \cite{ATLAS:2016cyf,CMS:2016abv} is about $m_{Z'}/g_{B-L}\gtrsim20$~TeV ($10$~TeV) for $m_{Z'}=4$~TeV ($5$~TeV)~\cite{Okada:2016gsh,Okada:2018ktp}, respectively. If the B-L charge of the dark matter is $n\sim\mathcal{O}(1)$, the direct detection constraints can be stronger than the ones from current collider experiments. Moreover, if we consider the relic abundance of the DM, which can be estimated as \cite{Duerr:2015wfa}
\beq
\Omega_\chi h^2&\approx&\frac{2.14\times10^9~\mathrm{GeV}^{-1}}{\langle\sigma v\rangle x_f^{-1}\sqrt{g_\ast}M_{Pl}}\nonumber\\
&\sim& 1.7\times10^{-4}\left[\left(1-\frac{4m_\chi^2}{m_{Z'}^2}\right)^2+\frac{\Gamma_{Z'}^2}{m_{Z'}^2}\right]\left(\frac{1~\mathrm{TeV}}{m_\chi}\right)^2\left(\frac{m_{Z'}/(\sqrt{n}g_{B-L})}{1~\mathrm{TeV}}\right)^4~,
\eeq
then, using the bound $m_{Z'}/(\sqrt{n}g_{B-L})\gtrsim20$~TeV from direct detection, we find that
\beq
\Omega_\chi h^2\gtrsim 27\times\left[\left(1-\frac{4m_\chi^2}{m_{Z'}^2}\right)^2+\frac{\Gamma_{Z'}^2}{m_{Z'}^2}\right]\left(\frac{1~\mathrm{TeV}}{m_\chi}\right)^2~.
\eeq
To be consistent with the current cosmological observation~\cite{Planck:2018vyg}, the DM mass must be very close to the resonance ($2m_\chi\approx m_{Z'}$), which is unnatural if this must occur simply as a coincidence.

The situation is very different if we consider that there is an extra $\textrm{U(1)}_X$ gauge symmetry and the cancellation mechanism is applied. The gauge symmetry is now $\textrm{SU(2)}_L\times \textrm{U(1)}_Y\times \textrm{U(1)}_{B-L}\times \textrm{U(1)}_X$. A Dirac fermion in the representation $\chi\sim(1,0,n_\chi n_\phi,n_\chi)$ is introduced to be the DM candidate, while two complex scalar fields in representation $\Phi_1\sim (1,0,1 ,0)$, $\Phi_2\sim (1,0,n_\phi,1)$ are also introduced to break the gauge symmetries. The Lagrangian is now given by
\beq
\mathcal{L}&\supset& -\frac{1}{4}C_{\mu\nu}C^{\mu\nu}-\frac{1}{4}X_{\mu\nu}X^{\mu\nu}+(D_\mu\Phi_1)^\dag D_\mu\Phi_1+(D_\mu\Phi_2)^\dag D_\mu\Phi_2+\bar{\chi}i\slashed{D}\chi,
\eeq
where the covariant derivatives are defined as,
\beq
D_\mu\Phi_1&=&(\partial_\mu-ig_{B-L}C_\mu)\Phi_1,\\
D_\mu\Phi_2&=&[\partial_\mu-i(n_\phi g_{B-L}C_\mu +g_XX_\mu)]\Phi_2,\\
D_\mu\chi&=&[\partial_\mu-in_\chi( n_\phi g_{B-L}C_\mu +g_XX_\mu)]\chi.
\eeq
Note that the charges of $\chi$ and $\Phi_2$ are chosen such that their couplings to the gauge fields are the same up to a factor $n_\chi$. It is worth emphasizing that this structure is one of the conditions for the cancellation mechanism.
Once $\Phi_1$ and $\Phi_2$ acquire non-zero VEV, $\langle\Phi_1\rangle=v_1/\sqrt{2}$ and $\langle\Phi_2\rangle=v_2/\sqrt{2}$, both $\textrm{U(1)}_{B-L}$ and $\textrm{U(1)}_X$ are spontaneously broken, and thus the gauge fields $C_\mu$ and $X_\mu$ become massive. The mass terms and the gauge interactions are given by
\beq
\mathcal{L}\supset\frac{g_{B-L}^2v_1^2}{2}C_\mu C^\mu+\frac{v_2^2}{2}(n_\phi g_{B-L}C_\mu+g_XX_\mu)^2+n_\chi\bar{\chi}\gamma^\mu\chi(n_\phi g_{B-L}C_\mu+g_XX_\mu)~.
\eeq
Now, as in the previous model, we can define a vector field, $V_\mu=X_\mu+(n_\phi g_{B-L}/g_X)C_\mu$, and rewrite the Lagrangian in terms of $V_\mu$ and $C_\mu$ as
\beq
\mathcal{L}&=&-\frac{1}{4}\left(1+\frac{n_\phi^2g_{B-L}^2}{g_X^2}\right)C_{\mu\nu}C^{\mu\nu}-\frac{1}{4}V_{\mu\nu}V^{\mu\nu}+\frac{n_\phi g_{B-L}}{2g_X}V_{\mu\nu}C^{\mu\nu}\nonumber\\
&&+\frac{g_{B-L}^2v_1^2}{2}C_\mu C^\mu+\frac{g_X^2v_2^2}{2}V_\mu V^\mu+n_\chi g_X\bar{\chi}\gamma^\mu\chi V_\mu~.
\eeq
Since the DM $\chi$ only couples to $V_\mu$, while SM fermions only couple to $W^a_\mu,~B_\mu$, and $C_\mu$, the only way for the dark sector to communicate with the SM is through the kinetic mixing between $V_\mu$ and $C_\mu$, which vanishes in the zero-momentum transfer limit. On the other hand, the annihilation cross section of $\chi$ is not suppressed, since it involves the s-channel processes and the total energy of the incoming dark matter pairs can be comparable to the vector masses. Now the observed relic abundance of DM can be satisfied without assuming that the mass of $\chi$ is near the resonance, because $m_{Z'}/(\sqrt{n_\chi n_\phi} g_{B-L})$ is no longer constrained by direct detection. Since the present work concerns the cancellation mechanism, we leave a detailed discussion of the phenomenologies of this model for future research.

\begin{figure}[tb]
	\centering
	\includegraphics[width=0.8\textwidth]{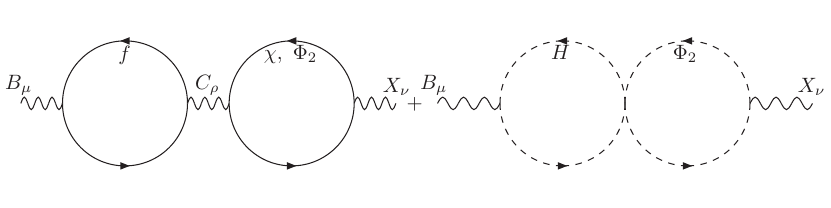}
	\caption{The 2-loop corrections for the $B_{\mu\nu}X^{\mu\nu}$ term.}
	\label{feyndiag5}
\end{figure}

Finally, we note that in principle, we should also include generic kinetic mixing terms among the $B_\mu,~C_\mu$ and $X_\mu$ fields. The kinetic mixing between $C_\mu$ and $X_\mu$ does not violate the cancellation mechanism because it only changes the coefficient of the kinetic terms for $C_\mu$ and $V_\mu$, which are unimportant for the direct detection. The $C_{\mu\nu}B^{\mu\nu}$ mixing does not violate the cancellation, to the leading order, since the DM only couples to $V_\mu$. The kinetic mixing term $B_{\mu\nu}X^{\mu\nu}$ can lead to a non-vanishing $\chi$-nucleon scattering cross section even in the limit $q^\mu\to0$. However, if we assume that the magnitudes of the kinetic mixing terms are of the same order as their leading loops corrections, the $B_{\mu\nu}X^{\mu\nu}$ term can be naturally small since its leading corrections come from 2-loop diagrams, as shown in FIG.\ref{feyndiag5}~. Even in other scenarios, in which some SM fields are charged under the $\textrm{U(1)}_X$ symmetry,\footnote{Note that the up and down quarks should not be $\textrm{U(1)}_X$ charged otherwise the cancellation is violated.} and one-loop corrections for $B_{\mu\nu}X^{\mu\nu}$ can be generated, the direct detection constraint can still be satisfied, since a mixing with $s_\epsilon\sim10^{-3}$ is small enough for a TeV-scale $V_\mu$ mediator.

\section{Conclusion}\label{concl.}
In this work, we discussed two methods for proving the cancellation mechanism for the Higgs-portal DM-quark scattering in the pNGB dark matter model. In the first proof, we used the non-linear representation of the complex singlet $S$, where the phase mode plays the role of the pNGB DM. We showed that the trilinear coupling between the on-shell pNGB DM $\chi$ and the radial mode $s$ is proportional to the momentum-squared of $s$ field, which vanishes in the limit of zero-momentum transfer. Since the $\chi$ field can only communicate with the quarks through the mixing between $s$ and the Higgs field $h$, the amplitude of the DM-quark scattering is suppressed by a small $t$ variable. We can easily generalize the model to inlcude the 2HDM or NHDM extension since all Higgs doublets can only couple to the pNGB $\chi$ through a mixing with the radial mode $s$. In addition, based on the non-linear representation we can easily generalize the model to the softly broken SO(N) cases and prove that the cancellation mechanism still works as well. We also found that in the SO(N) model, some soft-breaking cubic terms can be added without violating the cancellation mechanism for the DM.

In our second proof, we find that the combination of the CP-even scalars, which presents in the trilinear coupling with the pNGB DM, can be redefined as a new scalar field $\phi$ which does not couple to the SM fermions. In this picture, the masses of $h$ and $\phi$ are diagonalized while the kinetic terms of the Higgs bosons are not canonically normalized. A kinetic mixing between $\phi$ and $h$ also appears which is the only portal connecting the DM field to the SM fermions. Since the Higgs bosons only behave as mediators of the t-channel scattering, both their kinetic terms and mixing term vanish in the zero-momentum transfer limit, so that there is no communication between the dark and SM sectors to leading order.

Inspired by this second proof, we generalized the cancellation mechanism to include vector-portal models. In our first example, we considered a Dirac fermion electroweak doublet whose neutral component is the DM candidate. It is well known that the $Z_\mu$ boson mediated DM-nucleon scattering cross section is too large to accommodate the current direct detection bound. To solve this problem, we introduced a new gauged U(1)$_X$ symmetry and proved that the amplitude induced by the new gauge boson $X_\mu$ mediator can automatically cancel the amplitude induced by $Z_\mu$ boson. However, the cancellation is violated if a generic kinetic mixing term for the gauge bosons is included. The kinetic mixing parameter also needs to be fine-tuned in order to avoid the stringent direct detection bound.

In our second example, we considered a gauged $\textrm{U(1)}_{B-L}\times \textrm{U(1)}_X$ extension model and found that the cancellation mechanism works very well in this case. The kinetic mixing term which violates the cancellation is small if we assume that it has the same order of magnitude as its quantum correction, which is at the 2-loops level. Even if this is generated by 1-loop diagram, the resulting DM-nucleon cross section can still satisfy the current experimental bound since the vector mediator in this case can be as heavy as 1 TeV.

\acknowledgements
We would like to thank Zhao-Huan Yu and Yi-Lei Tang for discussions.
This work is supported by the National Natural Science Foundation of China (NSFC) under Grant Nos. 11875327 and 11905300, the China Postdoctoral Science Foundation under Grant No. 2018M643282, the Fundamental Research Funds for the Central Universities, the Natural Science Foundation of Guangdong Province, and the Sun Yat-Sen University Science Foundation.

\appendix
\section{Review of the cancellation mechanism for pNGB DM}\label{app1}
The couplings relevant to the $\chi$-nucleon scattering are the trilinear couplings as
\beq
\mathcal{L}_{h\chi^2}=\frac{1}{2}g_{h\chi^2}h\chi^2+\frac{1}{2}g_{s\chi^2}s\chi^2~,
\eeq
where $g_{h\chi^2}=2\lambda_{SH}v$, and $g_{s\chi^2}=2\lambda_{S}v_s$. We can rotate them in to the mass eigenstates of CP-even scalars using an orthogonal matrix, $O$, and write them in the following way:
\beq
\mathcal{L}_{h\chi^2}=\frac{1}{2}\chi^2(2\lambda_{SH}v,2\lambda_{S}v_s)O^T\begin{pmatrix}h_1\\ h_2\end{pmatrix}
\eeq
On the other hand, the Yukawa coupling between the Higgs and the SM fermions is given by
\beq
\mathcal{L}_{Yuk}=-\sum_f \frac{m_f}{v}h\bar{f}f=-\sum_f \frac{m_f}{v}\bar{f}f(h_1,h_2)O\begin{pmatrix}1\\ 0\end{pmatrix}~.
\eeq
Therefore, we can plot the t-channel Feynman diagrams (see FIG.\ref{fig1a}) for $\chi+q\to \chi+q$ scattering, mediated by $h_1$ and $h_2$, and compute the amplitude as
\beq
i\mathcal{M}=i(2\lambda_{SH}v,2\lambda_{S}v_s)O^T\begin{pmatrix}\frac{i}{t-m_1^2}&0\\0&\frac{i}{t-m_2^2}\end{pmatrix}O \begin{pmatrix}1\\ 0\end{pmatrix}\left(\frac{-im_q}{v}\right)\bar{u}_q(p_2)u_q(k_2)~.
\eeq
In the zero-momentum transfer limit ($t\to0$), it is easy to demonstrate that the amplitude vanishes. The proof given in Ref.~\cite{Gross:2017dan} is summarized as follows
\beq
i\mathcal{M}&\to& -i(2\lambda_{SH}v,2\lambda_{S}v_s)O^T\begin{pmatrix}\frac{1}{m_1^2}&0\\0&\frac{1}{m_2^2}\end{pmatrix}O \begin{pmatrix}1\\ 0\end{pmatrix}\left(\frac{m_q}{v}\right)\bar{u}_q(p_2)u_q(k_2)\nonumber\\
&=&-i(2\lambda_{SH}v,2\lambda_{S}v_s)O^T(M_{diag}^2)^{-1}O \begin{pmatrix}1\\ 0\end{pmatrix}\left(\frac{m_q}{v}\right)\bar{u}_q(p_2)u_q(k_2)\nonumber\\
&=&-i(2\lambda_{SH}v,2\lambda_{S}v_s)(M_{even}^2)^{-1} \begin{pmatrix}1\\ 0\end{pmatrix}\left(\frac{m_q}{v}\right)\bar{u}_q(p_2)u_q(k_2)\nonumber\\
&=&\frac{-i}{\mathrm{det}(M_{even}^2)}(2\lambda_{SH}v,2\lambda_{S}v_s)\begin{pmatrix}2\lambda_Sv_s^2&-2\lambda_{SH}vv_s\\-2\lambda_{SH}vv_s&2\lambda_Sv_s^2\end{pmatrix}\begin{pmatrix}1\\ 0\end{pmatrix}\nonumber\\
&&\times\left(\frac{m_q}{v}\right)\bar{u}_q(p_2)u_q(k_2)=0~.
\eeq
Therefore, the cross section of the $\chi$-nucleon scattering is suppressed by the very small momentum transfer of the cold dark matter.

\section{Vacuum configuration of the $K_{ijk}$ field}\label{app2}
Let $K_{ijk}$ ($i,j,k=1,2,3$) be a field in the traceless ($K_{ijk}\delta^{jk}=0$) symmetric 3-index representation of SO(3) symmetry. Since this representation is equivalent to the 7-plet of SU(2), we can relate $K_{ijk}$ field to the 7-plet real scalar field which has been studied in Ref.\cite{Cai:2015kpa}\footnote{In Ref.\cite{Cai:2015kpa}, the 7-plet of SU(2)$_L$ does not acquire any non-zero VEV, and the neutral component field is treated as a DM candidate.}. We denote the 7-plet field as $X_{\alpha_1\alpha_2\alpha_3\alpha_4\alpha_5\alpha_6}$ in this work (which is denoted as $\Phi_{ijklmn}$ in Ref.\cite{Cai:2015kpa}), where $\alpha_{1,2,3,4,5,6}=1,2$ are indices of the fundamental representation of SU(2), and they are totally symmetric by definition. The relation between $X$ and $K$ is given by
\beq
K^{ijk}=X_{\alpha_1\alpha_2\alpha_3\alpha_4\alpha_5\alpha_6}\varepsilon^{\alpha_1\beta}(\sigma^i)_{\beta}^{~\alpha^2}\varepsilon^{\alpha_3\gamma}(\sigma^j)_{\gamma}^{~\alpha^4}\varepsilon^{\alpha_5\delta}(\sigma^k)_{\delta}^{~\alpha^6}
\eeq
where $\varepsilon^{11}=\varepsilon^{22}=0,~\varepsilon^{12}=-\varepsilon^{21}=1$. The vacuum configuration that $\langle K_{112}\rangle=-\langle K_{222}\rangle\neq0$ and all other VEVs vanishing, can be translated into the VEVs of $X$ fields as
\beq\label{XVEV}
\langle X^{111111}\rangle=\langle X^{222222}\rangle\neq0~,
\eeq
and VEVs of all other components of $X$ fields are $0$. Another convenient notation of the 7-plet is the vector form, $X^I$ $(I=1,2,...,7)$, which relates to the tensor form as follows
\beq
X^{I}=\frac{1}{\sqrt{2}}\begin{pmatrix}\Delta^{(3)}\\\Delta^{(2)}\\\Delta^{(1)}\\\Delta^{(0)}\\\Delta^{(-1)}\\\Delta^{(-2)}\\\Delta^{(-3)}\end{pmatrix}=\frac{1}{\sqrt{2}}\begin{pmatrix}X^{111111}\\\sqrt{6}X^{111112}\\\sqrt{15}X^{111122}\\\sqrt{20}X^{111222}\\-\sqrt{15}X^{112222}\\\sqrt{6}X^{122222}\\-X^{222222}\end{pmatrix}~,
\eeq
where $(\Delta^{(Q)})^\ast=\Delta^{(-Q)}$. The vacuum eq.\eqref{XVEV} now becomes
\beq\label{DeltaVEV}
\langle \Delta^{(3)}\rangle=-\langle \Delta^{(-3)}\rangle\equiv -\frac{iv_\Delta}{\sqrt{2}}\neq0
\eeq
We should check whether this vacuum configuration can be obtained from the potential of $X$. The potential is given by
\beq
V_X=-\mu_X^2(X^\dag X)+\lambda_X(X^\dag X)^2+\frac{\lambda'_X}{48}(X^\dag S^{ij} X)(X^\dag S^{ij} X)~,
\eeq
where $S^{ij}\equiv \{T_7^i,T_7^j\}/2-4\delta^{ij}$, and $T_7^i$ ($i=1,2,3$) are generators of SU(2) for 7-plet (find more details of the generators in Ref.\cite{Cai:2015kpa}). The explicit expression of the $\lambda'_X$ term in the potential is a little bit lengthy, so we only present the terms involving at least quadratic of $\Delta^{(3)}$:
\beq
\frac{\lambda'_X}{48}(X^\dag S^{ij} X)(X^\dag S^{ij} X)&=& \lambda'_X\left[\frac{25}{32}\left|\Delta^{(3)}\right|^{4}\right.\nonumber\\
&&\left.-\left(\frac{5}{8}\left(\Delta^{(0)}\right)^{2}+\frac{5}{16}\left|\Delta^{(1)}\right|^{2}-\frac{25}{16}\left|\Delta^{(2)}\right|^{2}\right)\left|\Delta^{(3)}\right|^{2}+\dots\right]
\eeq
We assume the vacuum configuration eq.\eqref{DeltaVEV} and define $\Delta^{(3)}=[\phi_3-i(v_{\Delta}+a_3)]/\sqrt{2}$, then substitute it into the potential. Finally, we can find the minimum of the potential is given by
\beq
-\mu_X^2+\lambda_Xv_3^2+\frac{25}{32}\lambda'_Xv_3^2=0~,
\eeq
while the mass spectrum of the scalar fields are
\beq
&&m_{\phi_3}^2=m_{\Delta^{(2)}}^2=0~,\\
&&m_{a_3}^2=2\lambda_Xv_3^2,\qquad m_{\Delta^{(0)}}^2=-\frac{45}{32}\lambda'_Xv_3^2,\qquad m_{\Delta^{(1)}}^2=-\frac{15}{16}\lambda'_Xv_3^2~.
\eeq
We can see that there are three massless NGBs, which is due to fact that the SO(3) symmetry is completely broken by the vacuum. These NGBs can be absorbed by gauge fields if we gauge the SO(3) symmetry. The masses of $a_3$,~$\Delta^{(0)}$, and $\Delta^{(1)}$ fields imply that the vacuum is stable only when $\lambda_X>0$, and $\lambda'_X<0$. In conclusion, for some proper values of the parameters, the vacuum configuration that $\langle K_{112}\rangle=-\langle K_{222}\rangle \neq0$ can be automatically obtained.

\bibliographystyle{JHEP-2-2}
\bibliography{ref}

\end{document}